\documentstyle[12pt,epsf]{article}  
\newcommand{\rrr}{\rangle}

\newcommand{\HH}{{\cal H}}

\newcommand{\TT}{{\cal T}}

\newcommand{\wt}{\widetilde}

\newcommand{\be}{\begin{equation}}
\newcommand{\ee}{\end{equation}}
\newcommand{\ben}{\begin{eqnarray}\displaystyle}
\newcommand{\een}{\end{eqnarray}}
\newcommand{\refb}[1]{(\ref{#1})}

\begin{document}

{}~ \hfill\vbox{\hbox{hep-th/9912249}\hbox{MRI-PHY/P991238}
\hbox{CTP-MIT-2934}}\break

\vskip 3.5cm

\centerline{\large \bf Tachyon Condensation in String Field Theory}

\vspace*{6.0ex}
\centerline{\large \rm Ashoke Sen
\footnote{E-mail: asen@thwgs.cern.ch, sen@mri.ernet.in}}

\vspace*{1.5ex}

\centerline{\large \it Mehta Research Institute of Mathematics}
 \centerline{\large \it and Mathematical Physics}

\centerline{\large \it  Chhatnag Road, Jhoosi,
Allahabad 211019, INDIA}

\vspace*{1.5ex}

\centerline{and}

\vspace{1.5ex}

\centerline{\large \rm Barton Zwiebach
\footnote{E-mail: zwiebach@mitlns.mit.edu}}

\vspace*{1.5ex}

\centerline{\large \it Center for Theoretical Physics}
\centerline{\large \it Massachussetts Institute of Technology}
\centerline{\large \it  Cambridge, MA 02139, USA}

\vspace*{4.5ex}
\medskip
\centerline {\bf Abstract}

\bigskip
It has been conjectured that at a stationary point 
of the tachyon potential for 
the D-brane of bosonic string theory, 
the negative energy density 
exactly cancels the D-brane tension.
We evaluate this tachyon  potential by  
off-shell calculations in open
string field theory.
Surprisingly, the condensation of
the tachyon mode alone into the stationary point
of its cubic potential is found to 
cancel about 70\% of the D-brane tension. 
Keeping relevant scalars up to four mass levels
above the tachyon, the energy density 
at the shifted stationary point cancels  99\% of the 
D-brane tension.

\vfill \eject

\baselineskip=18pt

It has been argued on various general grounds that the classical
tachyon potential on a D-$p$ brane of the bosonic string theory has a
stationary point where the total negative potential energy due to the
tachyon
exactly cancels the tension of the D-brane\cite{9902105,RECK}. 
At this stationary point the configuration is 
indistinguishible from the vacuum where there is no
brane. A similar argument can be given for the tachyon potential 
on a D-brane anti-D-brane
system or a non-BPS D-brane of type II string
theories \cite{9805019,9805170,9808141,9810188,9812031,9812135}.
There is, however, no direct proof of these relations. 

In this paper we 
demonstrate this phenomenon directly using
string field theory. We restrict ourselves to bosonic string
theory and use Witten's string field theory with cubic
action \cite{WITTENSFT}, although in
principle the version of open string field theory suitable for 
off shell inclusion of closed strings \cite{openclosed} could 
have been used as well. The analysis could be performed  
for  superstring theories as well
using open superstring field theory\cite{WITTENSUSFT}. 
The background independent features of the tachyon potential
noted in \cite{9911116} make earlier studies of this
potential in string field theory 
\cite{KOST,KS,KOST2,9409015,9412106,9605088}\footnote{For 
studies of the tachyon potential in the first quantized 
formulation, see \cite{BERD,BANKS}.} relevant to the problem
of D-brane annihilation. 
Indeed, in a very interesting paper, 
Kostelecky and Samuel \cite{KS} gave evidence that the stationary
point of the cubic tachyon potential survives with controllable 
corrections the inclusion of higher
mass scalars of the string field expansion. Further evidence 
to this effect was given in \cite{9605088}. It is this
non-perturbative vacuum that we focus on in the present paper.
Our present advantage is that we have  
an explicit conjecture 
for the value of the potential at the 
stationary point we are looking for. Hence we can compare the
results obtained from string field theory with 
the conjectured value. As
we shall see, using a suitable approximation scheme, we can find 
a stationary point
of string field potential 
where  the value of the
potential is about 1\% away  from
 the conjectured answer.\footnote{Our
calculations appear to be consistent with the related
computations of \cite{KS,9605088}.} 

\noindent
$\underbar{The conjecture and the setup}$. 
Some general properties of the tachyon
potential in string field theory were analysed 
in \cite{9911116}, where it was shown
that the tachyon potential on
a D-brane of bosonic string theory takes a universal form:
\be \label{e1}
V(T) = M f(T) \, ,
\ee
where $M$ is the mass of the D-brane\footnote{We are assuming 
that all the
directions tangential to the brane are compact so that the brane has a
finite mass.} and $f(T)$ is a universal function independent of the
background in which the D-brane is embedded. The tachyon 
field $T$ and the
function
$f(T)$ are defined as follows\cite{9911116}. Let $\HH$ denote the 
space of states of ghost number one 
of the two-dimensional conformal field theory
of the
$(b,c)$ ghost system and a matter system of central charge 26 
on the upper
half plane, and let
$|0\rangle$ denote the SL(2,R) invariant vacuum of this conformal field
theory.
Let $\HH_1\subset \HH$ denote the space of states of ghost number one
obtained by acting on
$|0\rangle$
with  oscillators $b_n$, $c_n$, and  matter Virasoro generators $L_n$.
The subspace $\HH_1$ of $\HH$ is a background independent subspace
containing the zero momentum tachyon state $c_1|0\rangle$ and
having the property that 
we can consistently set the component of the string field along
$\HH-\HH_1$ to zero in looking for a solution of the equations of motion.
Since all fields in $\HH_1$ may acquire expectation
values, the  real problem is finding a stationary point
of the string field potential $V(T)$  associated to
the string  field $|T\rangle$ corresponding to a general state in
$\HH_1$. This string field $|T\rangle$, still called here the 
tachyon field,
includes an infinite collection of variables corresponding to the
coefficients of
expansion of a state in $\HH_1$ in some basis.\footnote{Although we shall
refer
to these coefficients as fields, we should keep in mind that these
represent zero
momentum modes 
of the fields corresponding to space-time independent field
configurations.}

\medskip
The function $f(T)$ is given by the following string field
theory 
expression: 
\be \label{e2}
f(T) = 2 \pi^2 \,\Big(\,\,{1 \over 2} \langle I\circ \TT(0) 
Q_B \TT(0) \rangle +
{1\over 3}
\langle h_1 \circ \TT(0) h_2 \circ 
\TT(0) h_3 \circ \TT(0) \rangle \,\Big)\, .
\ee  
Here  $Q_B$ is the BRST charge, and $\TT(z)$ 
denotes the two dimensional field
which creates the state
$|T\rrr$ from the SL(2,R) invariant vacuum:  
$|T\rangle = \TT(0)|0\rangle\, .$
$I$, $h_1$, $h_2$ and $h_3$ are a set of familiar conformal 
transformations\cite{LPP} 
whose expressions were reviewed in \cite{9911116}. 
Given any conformal transformation described by the function $h(z)$, and a
vertex operator $\Phi(z)$ of the conformal field theory,
$h\circ\Phi(0)$ denotes the conformal transform of $\Phi(0)$ by
$h$. Thus for example if $\Phi$ denotes a dimension $d$ primary field, 
then $h\circ\Phi(0)=(h'(0))^d\Phi(h(0))$. For non-primary fields there
will be extra terms involving higher derivatives of $h$. Finally $\langle
~ \rangle$ denotes the correlation function in the conformal field theory
of matter and ghost fields, normalized so that
$\langle c_{-1} c_0 c_1 \rangle = 1\,$.\footnote{Note that this
differs from the convention of ref.\cite{9911116} by a factor of $L$, $-$
the (infinite) length of the time interval. This is due to the fact that 
we are writing down the expression for the potential instead of the
action. We can make these two notations consistent 
by 
choosing $L=1$; in that case the potential can be identified to the
negative of
the action. The final results of course are independent 
of $L$.}$^,$\footnote{The factor of $2\pi^2$ in eq.\refb{e2} 
arises as follows.
With $S(\Phi) = -{1\over g_0^2} \Bigl({1\over 2} 
\langle \Phi, Q_B \Phi\rangle + \cdots \Bigr)$, the D-brane mass is : 
$M = 1/(2\pi^2 g_0^2)$\cite{9911116}. 
Then $V(T) = - S(T) =  M \cdot (2\pi^2) 
\Bigl({1\over 2} \langle T, Q_B T\rangle +
\cdots \Bigr) = M f(T).$} 

\medskip
The conjecture of ref.\cite{9902105} can now be restated as follows. In
the space $\HH_1$ there must be a state $|T_c\rangle$ such that $f(T)$
has
a stationary point at $T=T_c$, and
\be \label{e7}
f(T_c) = -1\, .
\ee
The total D-brane mass at $T=T_c$ vanishes:
$M + V(T_c) = M ( 1 + f(T_c))=0$.  

\bigskip
\noindent
$\underbar{Zeroth Approximation}$.
We proceed to verify this conjecture using a systematic approximation
scheme suggested
by Kostelecky and Samuel\cite{KS}. In order to explain this procedure, let
us first consider setting all components of $|T\rangle$ 
to zero except for the
coefficient of the state $c_1 |0\rangle$, a state that will be said
to be of {\it level zero}.  
Thus we take 
\be \label{e9} 
|T\rangle = t\, c_1 |0\rangle\, . 
\ee
Substituting this into eq.\refb{e2} we get the zeroth approximation to
the tachyon potential
\be \label{e10}
f^{(0)}(t) = 2 \pi^2 \Big(-{1\over 2} t^2 + {1 \over 3} 
\,  {t^3\over r^3} \,\Big)\,, 
\,  \qquad  r = {4\over 3 \sqrt 3} .
\ee
This has a local minimum at
\be \label{e11}
t = t_c \equiv r^3 = \Big({4 \over 3\sqrt 3}\Big)^3 \simeq  0.456\, .
\ee
At this minimum
\be \label{e12}
f(t_c) =- 2 \pi^2 \cdot {1\over 6} \cdot r^6
= -  {\pi^2 \over 3}
\cdot \Big({4 \over 3 \sqrt 3}
\Big)^6 =  - {4096\over 59049} \,\pi^2 
 \simeq -0.684\, . 
\ee
We found it very encouraging that this zeroth order approximation
to the vacuum energy at the stationary point gives essentially
70\% of the expected value!  In fact, 
the off-shell choice of 
cubic string field theory (as opposed to string field theory with
higher order vertices) yields
at this level the best possible approximation to the expected value.
Indeed, the constant $r$ defined above is essentially the mapping
radius of the disks defining the three string vertex \cite{9409015},
and it is maximal for the vertex of the 
cubic theory. Thus
$|f(t_c)|$ is maximal for this choice.

\medskip
\noindent
$\underbar{Subsidiary conditions on $T$}$.
In order to compute corrections to this result, we need to include the
higher level fields in our analysis. The analysis can be simplified by
noting
that the potential \refb{e2} has a twist symmetry under which all
coefficients of states
at odd levels above $c_1|0\rangle$ change sign, whereas
coefficients of states at even level
above $c_1|0\rangle$ remain unchanged \cite{KS,9705038}.\footnote{The
origin
of this symmetry can be traced to the relations 
$h_1(-z)=\wt I(h_3(z))$, 
$h_2(-z)=\wt I(h_2(z))$, and $h_3(-z)=\wt I(h_1(z))$, where $\wt
I(y)=1/y$ is a combination of the SL(2,R) and world-sheet parity
transformations. In the restricted sector $\HH_1$ the world-sheet parity 
as well as SL(2,R) is a symmetry of the theory.} Thus coefficients of
states at odd levels
above $c_1|0\rangle$ must always enter the action in pairs, and we can
trivially satisfy
the equations of motion of these fields by setting them to zero.
Thus we look for solutions where
$|T\rangle$ {\it contains only even level states}. With the
state $c_1|0\rangle$ defined to be at level zero, the  
additional fields we must consider will be at levels two, four, and higher.
At level two, for example,  
we find  three states, 
$c_{-1}|0\rangle$, $L_{-2}
c_{1}|0\rangle$ and $b_{-2}c_0c_1|0\rangle$.  

We can 
further simplify the expansion by using the Feynman-Siegel gauge:
\be \label{e16}
b_0 |T\rangle = 0\, .
\ee
This gauge choice can be justified by first showing that such a gauge can
be chosen at the linearized level, and then assuming that the fields are
small enough so that we can continue to make this gauge choice even in the
presence 
of interactions. The proof of validity of this gauge at the
linearized level proceeds
as follows. Let $|T^{(2n)}\rangle$ denote an arbitrary level $2n$ state in
$\HH_1$. Let us define
$|\Lambda^{(2n)}\rangle = b_0 |T^{(2n)}\rangle \, .$
Then $|\wt T^{(2n)}\rangle \equiv  |T^{(2n)}\rangle - (2n-1)^{-1} Q_B
|\Lambda^{(2n)}\rangle$ satisfies the desired gauge condition
$b_0 |\wt T^{(2n)}\rangle  = 0$.\footnote{In deriving the 
above equation we have used that (a)
$\{ Q_B, b_0\} = L_0^{tot}$, with $L_n^{tot}$ 
denoting the combined Virasoro generators of
the matter and the ghost sectors, 
(b) $b_0 |\Lambda^{(2n)}\rangle =0$ due
to the relation $(b_0)^2=0$, and that (c) 
the $L_0^{tot}$ eigenvalue of a
level $2n$ state is $(2n-1)$.} This shows that for $n\ge 1$, 
it is possible to gauge
transform a general level $2n$ state $|T^{(2n)}\rangle$ to a state $|\wt
T^{(2n)}\rangle$
satisfying the Feynman-Siegel gauge.

The equations of motion in the
Feynman-Siegel gauge are equivalent to the equations of motion of
 the gauge invariant action 
if there are no
residual gauge
transformations which act non-trivially and
preserve the gauge, 
{\it i.e.} if  
there are no pure gauge directions inside the subspace \refb{e16}. 
Assume  there is pure gauge direction
$|\eta^{(2n)}\rangle$ satisfying the  gauge condition 
\refb{e16}.
Then, 
$Q_B |\eta^{(2n)}\rangle = 0$,
and together with the gauge condition \refb{e16}, gives 
$L_0^{tot}|\eta^{(2n)}\rangle = \{Q_B, b_0\} |\eta^{(2n)}\rangle = 0$.
Since $L_0^{tot}|\eta^{(2n)}\rangle= (2n-1) |\eta^{(2n)}\rangle$, we see
that $|\eta^{(2n)}\rangle$ must vanish, as we wanted to show. 
Hence the Feynman-Siegel gauge is a valid gauge choice 
for sufficiently small field configurations.

\medskip
\noindent
$\underbar{Approximation with level two fields}$.
 Using the Feynman-Siegel gauge
we have
\be \label{e13}
|T\rangle = t c_1 |0\rangle + u c_{-1}|0\rangle + v \cdot {1\over
\sqrt{13}} L_{-2} c_1|0\rangle\, , 
\ee
which includes the two level two fields $u$ and $v$. 
The $(1/\sqrt{13})$ factor in the
 last term was  chosen for convenience ($L_n$'s denote 
matter Virasoro generators.).
At this stage we can simply substitute \refb{e13} into
\refb{e2} and find $f(t, u, v)$, but there is a further
approximation which is possible\cite{KS}. For this let us define the level
of a  
given term in $f(T)$ as the sum of the levels of all the fields
appearing in this term. We can now approximate
the potential $f(T)$ by keeping only terms up to a certain level. Since
the quadratic terms involving level two fields is already level four, it
does not make sense to truncate the potential to terms below level four once 
we have included the level two fields in our analysis. Thus
the next approximation to the potential, $f^{(4)}$, will be obtained by
substituting in \refb{e2} the expansion \refb{e13} and keeping only
$t^3$, $t^2u$, $t^2v$, $tu^2$, $tv^2$ and
$tuv$ interaction terms. Since we cannot have fields appear
in interactions before their quadratic 
terms appear, we define 
{\it the level $2n$ approximation $f^{(2n)}$ to contain
all interaction terms up to level $2n$ built from fields 
up to level $n$}. Thus at level six,
we do not  include any new fields (odd level fields are
set to zero) but we need to include four new interactions:
 $u^3$, $v^3$, $u^2v$ and
$uv^2$.

Ref.\cite{KS} evaluated the potential up to level six and found that the
stationary point 
of the potential persists up to this level. Their
result,
when translated into the normalization convention of this paper, is
as follows.\footnote{In order to convert the result of ref.\cite{KS} to
our convention, we need to set $\alpha'=1$ and $g=2$ in the expressions
given in ref.\cite{KS}, and multiply their
potential by a factor of $2 \pi^2$. The fields $t$, $u$ and $v$
used in our 
paper correspond to their fields $\phi$, $-\beta_1$ and $B$
respectively.} At  level four the potential is given  
by
\ben \label{e14} 
f^{(4)}(T) &=&  2 \pi^2 \Big( -{1\over 2} \,  t^2 +{3^3   
\sqrt{3}\over 2^{6}} \,
t^3
\nonumber \\
&&
-{1\over 2} \,u^2\,\,
+\,\, {1\over 2} \,\, v^2 
\,\, + { 11\cdot 3\sqrt{3} \over 2^6} \, t^2\, u
-{5\cdot 3\sqrt{39} \over\,2^6}\,  t^2 \,v \nonumber \\
&& 
 \,\, + {19\over 2^{6}\,\sqrt{3}}\,t\,u^2
+ {7\cdot\,83\over 2^{6} \cdot 3 \sqrt{3}}\,t\,v^2
-{ 11 \cdot 5\sqrt{13}\over 
2^5\cdot 3\sqrt{3}} t\,u\,v \Big)\, . 
\een
$f^{(4)}(T)$
has a stationary 
point\footnote{Strictly, this point is not a minimum  
of the potential, nor even a
local minimum. This is not  necessarily problematic. 
The string field theory has
ghost and auxiliary modes with negative mass squared 
that are not physical
tachyons. The field $u$ appearing 
in \refb{e14} is an example of this. We still expect 
physical stability of this stationary point.} 
at $T_c$ ($t_c\simeq 0.542$, $u_c\simeq 0.173$, 
$v_c\simeq 0.187$) 
at which $f(T_c)\simeq -0.949$. This is about 95\% of the expected answer
$-1$!\footnote{Since ref.\cite{KS} did not have a reference scale to
compare with, they expressed the potential in units of 
the string tension and
the on-shell
three tachyon coupling, and concluded that the potential 
is quite shallow.
On the other hand, using the mass of the D-brane as the reference scale,
we see that the potential is in fact quite deep. 
Already at this level it
is about 95\% of the mass of the D-brane.}

At level six  the potential includes the level four interactions plus
additional terms:
\ben \label{e15}
f^{(6)}  
\hskip-8pt &=\hskip-8pt & f^{(4)} 
+  2 \pi^2 \, \Big( {1\over 2^{6} \sqrt{3} }\,
\,u^3 
\, -\,  {7\cdot 41 \cdot 73 \over  3^4 \cdot  2^6\sqrt{39} }\,\,v^3
-\, {5\cdot 19 \sqrt{13} \over 2^6\,3^3 \sqrt{3} }\, \,u^2\,v 
   +\, { 11 \cdot 7\cdot 83\over 2^6 \cdot 3^4 \sqrt{3}}\,
\,u\,v^2 \Big) . 
\een
Solving the equations of motion that follows from 
 the total level six potential $f^{(6)}$, 
one finds that the 
location of $T_c$ is shifted slightly 
($t_c\simeq 0.544$, $u_c\simeq 0.190$, $v_c\simeq 0.202$) with 
$f(T_c)\simeq -0.959$. By including two modes in addition to the
tachyon we have gone from 68\% to 96\% of the expected vacuum
energy! This is certainly encouraging and
leads us to believe that
the expansion converges rapidly to the expected answer.

\medskip\noindent 
$\underbar{Approximation with level eight interactions}$.
To establish the convergence beyond reasonable
doubt,
we now undertake the substantially more involved calculation
of the potential to level eight. 
For this  we need to include  all the level four fields. 
A general tachyon field configuration in the Feynman-Siegel gauge,
including fields up to level four, has the form:\footnote{Since we
are using background independent modes, we have here less 
fields than in Refs.\cite{KS,9605088}, the last of which 
cites a computation of the potential using level six fields.}
\ben \label{e21}
|T\rangle &=& t c_1 |0\rangle + u c_{-1}|0\rangle + v \cdot
{1\over
\sqrt{13}} L_{-2} c_1|0\rangle \nonumber \\
&& + A L_{-4}c_1 |0\rangle + B L_{-2} L_{-2} c_1 |0\rangle + C c_{-3}
|0\rangle \nonumber \\
&& + D b_{-3} c_{-1} c_1 |0\rangle + E b_{-2} c_{-2}
c_1 |0\rangle
+ F L_{-2} c_{-1} |0\rangle \, .
\een
In order to construct the potential to level eight, we need to substitute
\refb{e21} into \refb{e2}, and evaluate the action keeping terms up to
level eight. We use two different methods to compute the cubic interaction
vertices. In the first approach we explicitly compute the conformal
transformation of all the vertex operators associated with the state
\refb{e21} under the conformal maps $h_1$, $h_2$ and $h_3$, and compute
the three point correlation functions of the resulting operators. In the
second approach we use a representation of the matter Virasoro algebra in
terms of 26 free bosonic fields, and use the Neumann function method to
compute the three string vertex \cite{gross}. Both approaches give the
same results.

Besides the terms given in
eqs.\refb{e14} and \refb{e15} (which we explicitly verify), there are
four different kinds of additional terms 
in the computation of $f^{(8)}$. These are
\begin{enumerate}
\item The quadratic term involving the level 4 fields. This is a
level 8 contribution to the potential, and is given by:
\be \label{e22} 
\Delta_0 f^{(8)} = 2 \pi^2\,\, \Bigl( \,\,195 \, A^2 
+ \, 663 \, B^2\, +\, 234
\, AB \, 
 +\, 3 \, C D \, - \, {3\over 2} \, E^2 \, - \,
{39 \over 2} \, F^2\,\Bigr) \,\, .
\ee

\item There are new level four interaction terms, originating from the
coupling between two level 0 and a level 4 field. These are 
given by, 
\be \label{e22a}
\Delta f^{(4)} = 2 \pi^2 {1 \over \sqrt 3} \, t^2 \,\Big( \,{585 \over 32} 
A + {3523 \over 96}
B - {5 \over 12} C + {5\over 4} D + {19 \over 64} E
- {715\over 192} F \Big)\, .
\ee

\item There are new level six interaction terms, originating from the
coupling between a level 0, a level 2 and a level 4 field. 
These are given by 
\ben \label{e23}
\Delta f^{(6)}\hskip-7pt &=& \hskip-4pt  {2 \pi^2 \over \sqrt 3} 
\, t \,\Big[
\,\,u\,\Big(\, {715 \over 48}  A
+ { 38753 \over
1296 } B - {25 \over 54 } C + {25 \over
18 } D + {3827 \over 2592 } E 
 - {1235\over 864 } F \Big) \cr \nonumber
\\ && \hskip-9pt
+  \sqrt{13}\, v \Big( - {7495\over 1296 } A - {12101
\over 432 } B + {25 \over 162 } C  - {25 \over 54 } D
- {95  \over
864 } E + {6391 \over 2592 } F \Big) \Big]\, . 
\een

\item There are cubic interaction terms of level 8. These involve coupling
of
two level 2 fields with a level 4 field, and also the coupling of two
level 4 fields with a level 0 field. The 2-2-4 interaction 
terms are 
given by,
\ben \label{e24a}
\Delta_1 f^{(8)}\hskip-8pt &=& 
\hskip-6pt {2 \pi^2\over \sqrt 3} \Big[ u^2\Bigl( {1235
\over 864} A
 + {66937 \over 23328} B  - {5 \over 108} C  + {5 \over
36} D   + {124849 \over 139968} E - {65 \over
576} F \Bigr) \cr \nonumber \\
&& 
\hskip-15pt+\sqrt{13}\,uv
\Bigl( - {82445  \over 34992} A  - {133111  \over
11664} B  + {125  \over 1458} C - {125
 \over 486} D  - {19135  \over 69984} E 
 + {11039 \over 23328} F \Bigr) \cr \nonumber
\\
&& 
\hskip-15pt  +\, v^2 \,\Bigl( \, {254381 \over 23328} A
 + {1598597 \over 23328} B  - {2905
\over 8748} C  + {2905 \over 2916} D  + {11039 \over 46656} E
 - {230461 \over 46656} F \Big)\Big]\, .\nonumber \\
\een
On the other hand, the 0-4-4 interaction terms are given by,
\ben \label{e24}
\Delta_2 f^{(8)}\hskip-6pt &=&\hskip-6pt  {2 \pi^2\over \sqrt 3}\, t
\,\Big(\,{3539315 \over 11664} A^2 + {9440977 \over 17496} A B 
+ {4367233 \over 3888} B^2 - {325 \over 81} A C - {17615 \over 2187} B C 
\cr \nonumber \\ && + \,{25 \over
729} C^2 + {325 \over 27} A D + {17615 \over 729} B D +
{1598 \over 2187} C D  + {25 \over 81} D^2 + {1235 \over 432} A E 
\cr \nonumber
\\
&& 
+ \,{66937
\over 11664} B E + {665 \over 1458} C E - {665 \over 486} D E - {4061
\over 5184} E^2 - {1071785 \over 34992} A F
\cr \nonumber \\ &&  
 -\, {1730443 \over 11664} B F
+ {1625 \over 1458} C F - {1625 \over 486} D F 
- {248755 \over 69984} E
F + {143507 \over 46656} F^2\Big)\, .
\een
\end{enumerate}
The level 8 approximation to the full tachyon potential is obtained by
combining the contributions \refb{e14}, \refb{e15} and
\refb{e22}-\refb{e24}:
\be \label{e25}
f^{(8)} =  f^{(6)} +
\Delta f^{(4)} + \Delta f^{(6)} + \Delta_0 f^{(8)}  
+ \Delta_1 f^{(8)} + \Delta_2 f^{(8)}\, .
\ee

Given this potential we can search for 
a stationary point of the  potential.\footnote{The actual 
computations require use of symbolic
manipulation programs Maple and Mathematica.}  We find
that the equations of motion following from 
this potential are satisfied for\footnote{This stationary 
point 
is closest to the one given earlier for level four and level six
potentials,
in the sense that if we look for a numerical solution with the stationary
point of the level four or level six potential as the starting point, we
arrive at the solution \refb{e26}. There may be other solutions to the
equations of motion whose interpretation is not clear.}: 
\ben \label{e26}
&& t_c = 0.5482, \quad u_c = 0.2043, \quad v_c = 0.2045, \quad A_c =
-0.00495,
\quad
B_c = -0.00056, \nonumber \\
&& \quad C_c = -0.0549, \quad D_c = 0.0183, \quad E_c = 0.0317,
\quad F_c =
-0.0066\, .
\een
The value of the potential at this stationary point is given by 
\be 
f(T_c) = -0.9864\, .
\ee
This is about $1\%$ away from the expected answer $-1$. 
Note the near equality of the values of $u_c$ and $v_c$. We suspect
this to be an exact equality in the complete answer. This, in turn,
indicates that there might be a closed form expression for the state
$|T_c\rangle$ describing the stationary point, since otherwise it will be
very difficult to explain the equality of these coefficients.

\medskip
\noindent $\underbar{Discussion}$.
We shall end this paper by discussing 
the significance of our results and
some further investigations which these results suggest.
\begin{enumerate}

\item Our result indicates that the tachyonic vacuum of the
bosonic D-brane, representing  its annihilation, 
is described by a string field
dominated by the low lying modes of the theory. 
This is certainly 
surprising 
since total brane disappearance 
is a highly non-perturbative phenomenon 
and one could  have expected 
non-trivial participation of all the
higher string modes. 
As we saw, however, condensation of states up to level four 
account for 99\% of the potential energy 
required to cancel the tension of the D-brane. Our results
show that string field theory captures non-perturbative 
string dynamics. The string field $|T_c>$ appears 
to be well-defined.

\item Associated with the phenomenon of tachyon condensation on the
D-brane is the problem of the extra U(1) gauge
field \cite{9807138,9810188,9901159}. 
How does the  U(1) gauge field living on the brane disappear
after the tachyon condenses and the brane annihilates? 
Since the tachyon is
neutral, the gauge field cannot acquire mass via the Higgs
mechanism. 
Also, how do open strings with one endpoint  lying
on the D-brane in question disappear after tachyon condensation? 

Ref.\cite{9909062} proposed 
that at the extrema $T=T_c$, the action of the
gauge field vanishes identically. This explains
the absence of a dynamical gauge field. 
In addition, the path integral over the gauge field
now sets to zero the charged currents, 
thus explaining the absence of open strings 
with one endpoint on the brane in question.

Since our analysis shows that the tachyonic vacuum can be studied
efficiently with string field theory, one can ask if the above proposal
can also be verified using string field theory. In other words,
can one study the fate of action involving the gauge fields at the
extremum $T=T_c$ and show that its coefficient becomes small?

\item According to the conjecture of ref.\cite{9902105,RECK}, a D$(p-1)$
brane of the bosonic string theory can be regarded as a lump solution on a
D$p$ brane, where far away from the core of the lump the tachyon condenses
to the critical value $T_c$. Since the configuration $T=T_c$ seems to have
a good description in string field theory, it is natural to ask whether
the lump also has a good description in string field theory.

\item 
Another question that arises from our analysis is: is it
possible to write down a closed form
expression for the exact extremum $|T_c\rangle$ 
of the tachyon potential,
and/or of the lump solutions describing lower dimensional 
branes? 
As we have already mentioned, the near equality of 
$u_c$ and $v_c$
in eq.\refb{e26} can be taken as an evidence that there is a closed form
expression for
$|T_c\rangle$.

\item  Finally, we can wonder about the existence of a   
stationary point in the tachyon potential  for the bosonic
closed string field theory \cite{csft}. Could this vacuum, if
it exists, be  a state of unbroken general coordinate 
invariance having no dynamical graviton?  
While there appears to be no physical prediction for 
such hypothetical stationary state, the methods discussed
here may improve on earlier computations
\cite{KOST2,9409015,9412106} to 
give some new insight into this problem.  

\end{enumerate}

{\bf Acknowledgement}: A.S. would like to thank the
theoretical physics group at the Tata Institute of Fundamental Research 
for hospitality during the course of the work.

\end{document}